\documentclass[epj,nopacs]{svjour}
\pdfoutput=1
\usepackage[utf8]{inputenc}
\usepackage{caption,amsmath,amssymb,amsfonts,graphicx,cite,multirow,tikz,enumitem}
\DeclareCaptionType{Diagram}
\usepackage{pstricks,pdflscape,morefloats,algorithmicx,algpseudocode,slashed}
\usepackage[section]{algorithm}

\graphicspath{{graphics/}}
\makeatletter
\ifx\input@path\@undefined
\def\input@path{{graphics/}}
\else
\g@addto@macro\input@path{{graphics/}}
\fi
\makeatother

\usepackage{color}
\usepackage{hyperref}

\def\MSbar{$\overline{{\rm MS}}$}
\def\MSbareq{\overline{{\rm MS}}}

\usepackage{mathtools,feynman}

\preprint{HERWIG-2018-01\\ LU-TP-18-01\\  MCnet-18-01}

\title{Colour Rearrangement for Dipole Showers}

\author{Johannes Bellm\inst{1}}
\mail{johannes.bellm@thep.lu.se}
\institute{ Theoretical Particle Physics, Department of Astronomy and Theoretical Physics, 
Lund University, Lund, Sweden}

\date{Jan 18, 2018}

\abstract{ We present an algorithm to rearrange the colour chains of dipole
showers in the shower process according to the colour amplitudes of a simple matrix element. 
We implement the procedure in the dipole shower of Herwig and show comparisons to data. 
\PACS{{xx.yy.zz}{Xx Yy Zz}} }

\begin{document}

\maketitle

\section{Introduction}
\vspace{-3mm}
One of the main ingredients to describe data at collider energies in event generators 
like \cite{Herwig,Herwig7,Pythia8.2,Sherpa,Lonnblad:1992tz} are parton showers. 
In a probabilistic picture emissions off partons are produced according to probability 
distributions derived from the collinear and/or soft limit of matrix elements.

Aside from angular ordered shower approximations\cite{Marchesini:1987cf} so-called dipole 
showers\cite{Gustafson:1987rq} are implemented, as the 
divergent structure of QCD amplitude can be reproduced  in dipole-like emissions without 
double counting the soft wide-angle emissions. 
Prominent examples are e.g. the Ariadne shower\cite{Lonnblad:1992tz} or parton showers 
based on Catani Seymour (CS) dipole subtraction \cite{Catani:1996vz} earlier introduced for 
numerical calculation  of NLO corrections. CS showers are implemented in \cite{Dinsdale:2007mf},
\cite{Schumann:2007mg} and  \cite{Platzer:2009jq}. 
Further developments have been introduced in \cite{Hoche:2015sya} with 
the recent inclusion of collinear parts of the trilinear NLO splitting in 
\cite{Hoche:2017hno,Hoche:2017iem} and for antenna showers in \cite{Li:2016yez}.
Various methods have been introduced to correct the showering process with matrix 
element corrections at LO and NLO. These will not be discussed here. 
In this paper, we concentrate on the colour assignment of this kind of parton shower 
algorithms and implement an algorithm to rearrange the colour structure in the showering process.
Methods to correct the shower emissions with subleading $N_C$ effects have 
been discussed in \cite{Platzer:2012np} and \cite{Nagy:2015hwa}.
In this article, we discuss the possibility of shower emissions resulting in unfavoured 
colour configurations and introduce an algorithm to correct with simple LO matrix elements.
 
It is well known that QCD amplitudes can be decomposed into colour structures 
\cite{Mangano:1987xk}. Further colour bases and simplifications have been developed 
to calculate multileg amplitudes efficiently \cite{Maltoni:2002mq,Sjodahl:2014opa}. 
For event generation bases like Colourflow or Trace bases have the nice 
feature, that in the large $N_C$ limit ($N_C$ being the number of colours) 
the flows or traces can be interpreted as dipole chains/colour lines/strings. 
The starting conditions of a parton shower process and later the final conditions 
of the hadronisation process depend on the assignment of colour structures 
\cite{Ellis:1986bv,Maltoni:2002mq,Buckley:2011ms}. Event generators keep record  
of these colour lines and form either clusters of colour connected partons  
\cite{Herwig,Herwig7,Sherpa} or the lines themselves are interpreted as strings 
as in \cite{Pythia8.2}. 

The maximally helicity violating amplitude with one quark anti-quark pair ($q\bar{q}$) and 
$n$-gluons can  be written as  \cite{Mangano:1988kk},

\vspace{-6mm}
\begin{equation}
\sum_{col.} \| A_{1 ... n}\|^2 = \mathcal{F} \sum_{\{1,2...n\}} \frac{1}{(p_q1)(12)...(np_{\bar{q}})} 
+ \frac{1}{N_C^2} (interf.) \label{ColAmp}
\end{equation}
where $\mathcal{F}$ is a kinematic factor 
and $(i j)=2p_i\cdot p_j$ are products of the momenta of gluons and quarks. More general helicity structures are more
complicated. The sum on the right-hand side of Eq.~\ref{ColAmp} includes all permutations 
of the n-gluons and compared to QED the interference is suppressed by a factor of $N_C^{-2}$. 
With the suppression of interference terms the dominating term in the sum of Eq.~\ref{ColAmp} 
is the permutation with minimised numerator. In the string picture, this is the shortest string. 
This chain of dipoles can then be interpreted to have a stringlike behaviour\cite{Mangano:1988kk}. 
\vspace{-6mm}
\section{Dipole like showers}
\label{sec:dipoleshowers}
\vspace{-3mm}
In this section we summarize the algorithm implemented in Herwig 
\cite{Platzer:2009jq,Platzer:2011bc} emphasising the colour chain structure 
and allowing our language to cover other dipole-like shower and hadronisation procedures. 

\textbf{\underline{Starting the Shower:}} In dipole like showers usually, colour chains are assigned
 before the shower is allowed to evolve the event. A chain contains either one 
$q\bar{q}$-pair and $n$-gluons or only gluons. A chain is then given 
by $q - g_1-g_2-...-g_n-\bar{q}$, where a gluon contains both
 colour and anti-colour and is, therefore, a member of two dipoles. In the case with only two 
 gluons -- e.g. higgs production -- there are two colour dipoles between the gluons. In 
addition to the assignment of colour another initial condition is the chosen starting scale. 

\textbf{\underline{While Showering:}} These chains can then radiate gluons or break 
by a gluon splitting into a quark anti-quark state,  $g\to q\bar{q}$. This is done 
in a competing Sudakov veto algorithm. Here all dipoles of the chain produce an evolution scale 
and the dipole with the largest evolution scale is allowed to emit. The emission creates recoils
on the other participants in the chain, either because the spectator used to absorb the 
recoil is again connected to another dipole or the emitter can have another 
colour connection to another parton in the chain. 
If, in our example chain above, $q$ radiates with spectator $g_1$ the dipole spanned by
$g_1$ and  $g_2$ is modified as well. Thus, an emission of a gluon will affect 
the kinematics of up to four of the resulting dipoles.
If a gluon splits into a $q\bar{q}$-pair the colour of the dipole chain breaks and the quark 
carries the colour of the split gluon and the anti-quark gets the anti-colour.
Once the emission is performed the chain or two chains is/are evolved further until the shower 
algorithm is terminated by finding no emission scale above the infrared (IR) cutoff.  

\textbf{\underline{Splitting a Dipole:}}
Once a dipole in the chain is identified to be the winning participant, a momentum 
fraction $z$  is chosen according to the functional form of the splitting function. 
If spin-averaged splitting functions are used, the radiation angle $\phi$ of the emission 
plane around the dipole axis is chosen randomly on the interval $[0,2\pi)$. 
For spin-dependent splitting functions, the radiation angle can be biased by the 
helicity of the emitting parton. 
 
\textbf{\underline{After the Shower:}} Once the IR cutoff is reached the colour chains 
are interpreted as colour strings and hardonised in a Lund string model or the 
remaining gluons are split to break the chains in colour anti-colour $q\bar{q}$-pairs 
which then build the clusters of the cluster models. After forming of clusters/strings 
the process of colour reconnection can rearrange the constituents of the clusters/strings 
and in addition cluster fissioning or string breaking happens before cluster  masses/string 
lengths are reached that allows the conversion to hadronic states. 
  
\begin{Diagram}
\centering
\scalebox{0.3}{
\begin{feynman}
\fermion[]{0.5,0}{3,2}
\fermion[]{3,-2}{0.5,0}
\electroweak[]{0,0}{0.5,0}
\gluon[]{0.5,0}{1.5,0}
\gluon[]{1.5,0}{3,.7}
\gluon[]{1.5,0}{3,-.7}
\draw[red,line width=1.25mm] (3.5,2)   -- (4,.7) ;
\draw[blue,line width=1.25mm] (4,-0.7)  -- (4,.7) ;
\draw[green,line width=1.25mm] (3.5,-2) -- (4,-.7) ;
\draw[red,line width=1.25mm] (4.5,2)  -- (5,-.7) ;
\draw[blue,line width=1.25mm] (5,-0.7)  -- (5,.7) ;
\draw[green,line width=1.25mm] (4.5,-2)  -- (5,.7) ;
\end{feynman}
}
\caption{Simple process with three colour dipoles. This process is used to calculate 
the weight for the colour rearrangement.\label{fig:simpleDiagram}}
\end{Diagram}
\vspace{-6mm}
\section{Colour Rearrangement}
\vspace{-3mm}
If we interpret terms as in  Eq.~\ref{ColAmp} as probabilities to choose colour lines 
for the starting conditions of the showering process, we now want to know what 
happens to the colour structure after emitting off a dipole in a given dipole chain.
Emitting a gluon $g_a$ from dipole $g_1 - g_2$ \footnote{The following argument 
holds for emissions off the ends of the chains with a less complex structure.} 
leads us to:
   
   \vspace{-2mm}
  \begin{eqnarray}
  q -g_1 = g_2 - g_3 - g_4 -....- g_n - \bar{q} \label{CFS}\\
   (a)\rightarrow  q - g_1' - g_a - g_2' - g_3 - g_4 -....- g_n - \bar{q} \nonumber\\
   (b)\rightarrow  q - g_a - g_1' - g_2' - g_3 - g_4 -....- g_n - \bar{q} \nonumber\\
   (c)\rightarrow  q - g_1' - g_2' - g_a - g_3 - g_4 -....- g_n - \bar{q} \nonumber 
  \end{eqnarray}
   \vspace{-2mm}  
 
Here gluon $g_a$ should be identified as the softer gluon in the splitting of $g_1$ 
with a spectator $g_2$. Configuration (a) is obtained if the emission angle is such 
that the softer gluon is 'in between' the new $g_1'$ and $g_2'$. (b) is obtained 
when $g_a$ is closer to the quark $q_p$  than $g_1$. (c) is a configuration that is 
not obtained by CS showers but can happen if the emitter-spectator relation 
is not clear as in Ariadne. In the colourflow picture (a), (b) and (c) in Eq.~\ref{CFS} 
correspond to permutations of inner gluons and the weights of assigning the 
colour lines depends on the full chain. If we would  start the shower from the 
configuration received after emission we would assign the colours according 
to the weights in the colour representation. Here the emitted gluon 'feels' the 
nearby gluon and the colours are arranged accordingly. The independent dipole 
in a chain has no possibility to distinguish a preferred direction in terms of 
colour amplitudes.
 
In the physical picture where most of the emission of $g_1$ is in the angle 
opened by $g_1-g_2$ or possibly but suppressed closer (in this example) to the 
quark we can assume a shielding of colours  of dipoles $g_2' - g_3 - g_4 -...$.
Then these distant dipoles have little effect on the emission off $g_1$. The 
colour connected quark $q$, however, is close to the colour line of gluon $g_2'$.
In order to construct a weight to distinguish between configurations (a) and (b) 
we can  use the simplest matrix element available that includes three dipoles namely 
e.g. $\gamma* \to u \bar{u} g g$, see Diagram \ref{fig:simpleDiagram}. We are 
only interested in the distinction between colour structure $q - g_1' - g_a - g_2' - $ 
and $q - g_a - g_1' - g_2' - $ as the rest of the event remains unchanged. 
Even the identification of the $\bar{u}$ to represent gluon $g'_2$ is a good 
approximation as its colour charge vanishes in the weight ratio used to decide 
between the states.
 
In the actual implementation, we define a phase space point $\Phi$ from three dipoles and incoming beams to deliver the energy needed for the dipole combination.  
We use MadGraph to generate the process  $e^+e^- \to u \bar{u} g g$ and calculate 
the weights of  the squared colour amplitudes $w(1;2;3;4)$ = ${\tt jamp2[0]}$ and  
$w(1;3;2;4)$ = ${\tt jamp2[1]}$ at the given $\Phi$. If a flat random number in $[0,1)$ 
is smaller than 

\vspace{-2mm}
$$P_{\text{swap}}=\frac{w(1;3;2;4)}{w(1;2;3;4)+w(1;3;2;4)}$$ 
we swap the momenta of the gluons $g_2$ and $g_3$ which corresponds to rearranging the colour structure. Note that the weights take into account interferences and parts of off-shell effects but neglect the non-diagonal elements in the colour basis.

As the matrix element is simple and fast to compute we also allow swapping in the chain that was not modified by the last emission, by simply calculating the swapping probability for all neighboring triple dipoles. If the colour chain is already in an order preferred by the matrix element this will not change the probability of having this colour structure, see third comment in Sec.~\ref{sec:comments}, if not the lines will be rearranged to the 'preferred' order. Preferred is a probabilistic mixture of short or long chains which is now given by$P_{\text{swap}}$ rather than an uncontrolled function of evolution variable and emission angle.

\newpage
\section{Comments}
\label{sec:comments}
\vspace{-3mm}
We would like to add some comments on the rearrangements:
\begin{itemize}
\item By swapping to the preferred smaller dipole masses we allow fewer emissions in the following, 
as the dipole phase space is given by the mass of the dipole. 
\item Without proof, we assume that the coherence properties are  improved by allowing
to rearrange the colours such that the wide-angle emissions happen first, 
producing an ordered emission spectrum and therefore reduced chain length.
\item Assume a shower produces, for a given phase space point, the preferred colour structure 
with ME weight $w_1=w(1;2;3;4)$ with probability  $a$ and the unfavoured colour structure 
with probability $1-a$.
Then a swapping will produce the favoured colour structure as \footnote{Produce favoured 
and remain and produce unfavoured but change.},

\vspace{-5mm}
$$\left(1-\frac{w_2}{w_1+w_2}\right)\cdot  a +  
\left(\frac{w_1}{w_1+w_2}\right)\cdot (1-a)=\frac{w_1}{w_1+w_2}$$
\vspace{-4mm}

and similar for the unfavoured colour structure.
\item This method does not need weighted events to correct for the colour assignment. 
\item The rearrangement can be performed at any step of
 the shower and is not restricted to a given multiplicity. 
\item A possible failure of the method is the rearrangement to produce dipoles with 
masses that are too small to create colour singlets that further can decay to mesonic 
states. We did not yet observe this behaviour.
\item It is anticipated that we can use the same process to rearrange the colours of incoming partons if we do not allow the swapping of final state to initial state momenta. 
To do so we will in a further publication invert the incoming three-momenta and define all dipole participants as outgoing. As we sum over all helicity combinations this should give the correct weights.
\item Once the method is extended to LHC physics the colour reconnection model needs to be
reviewed/retuned as the rearrangement will create another density of cluster masses/strings sizes.
\item Using matrix elements with longer dipole chains e.g. $\gamma* \to u \bar{u} g g g$ 
to distinguish more permutations of intermediate gluons is part of future work.
\item It is clear that the method can be applied to any kind of dipole like shower e.g. the Sherpa
\cite{Schumann:2007mg}, the final state shower of Pythia \cite{Pythia8.2} as well as the Dire shower
\cite{Hoche:2015sya}. 
\end{itemize}
\vspace{-6mm}
\section{Results}
\label{sec:Results}
\begin{figure}[t]
\centering
\scalebox{.735}{\includegraphics{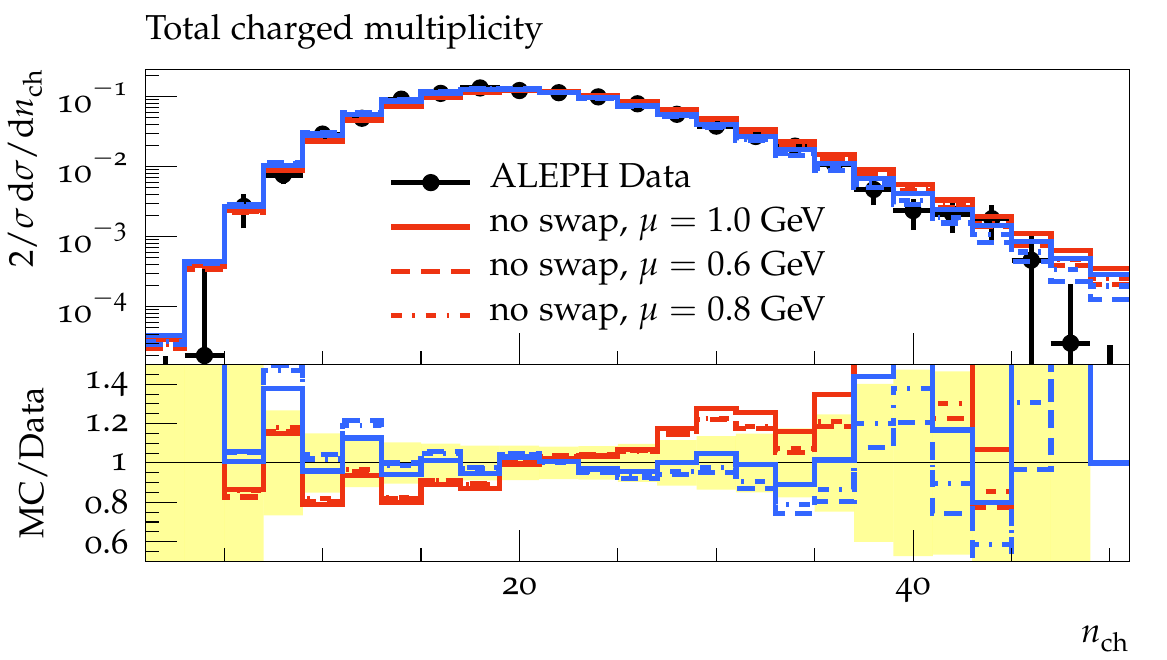}}
\scalebox{.735}{\includegraphics{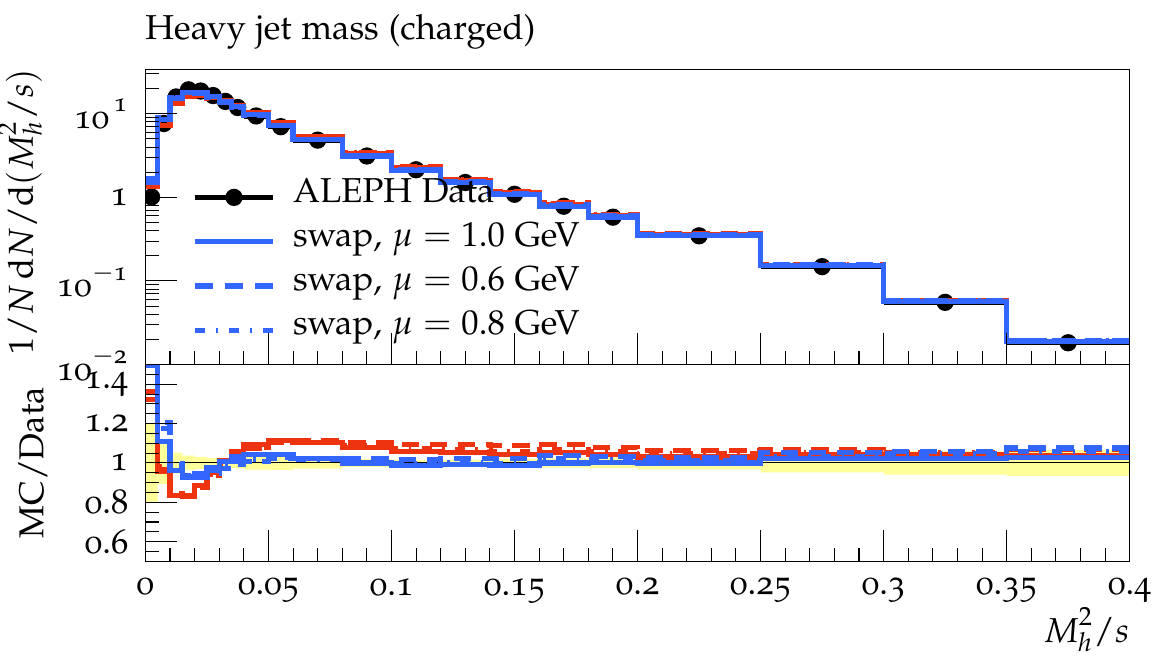}}
\caption{Upper plot: The total charge multiplicity as measured by \cite{Decamp:1991uz}. 
Lower plot: The heavy jet mass measured by \cite{Barate:1996fi}. For simulation 
setup see Sec.~\ref{sec:Results}  }
  \label{fig:LEP2}
\end{figure}

\begin{figure}[t]
  \centering
\scalebox{.735}{\includegraphics{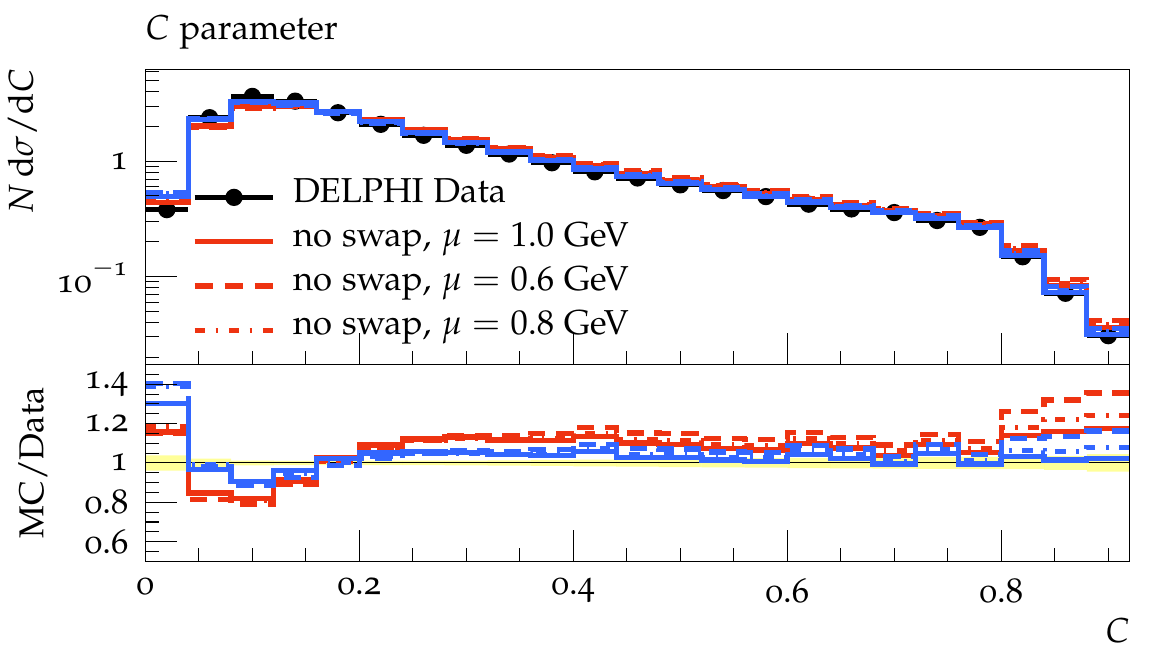}}
\scalebox{.735}{\includegraphics{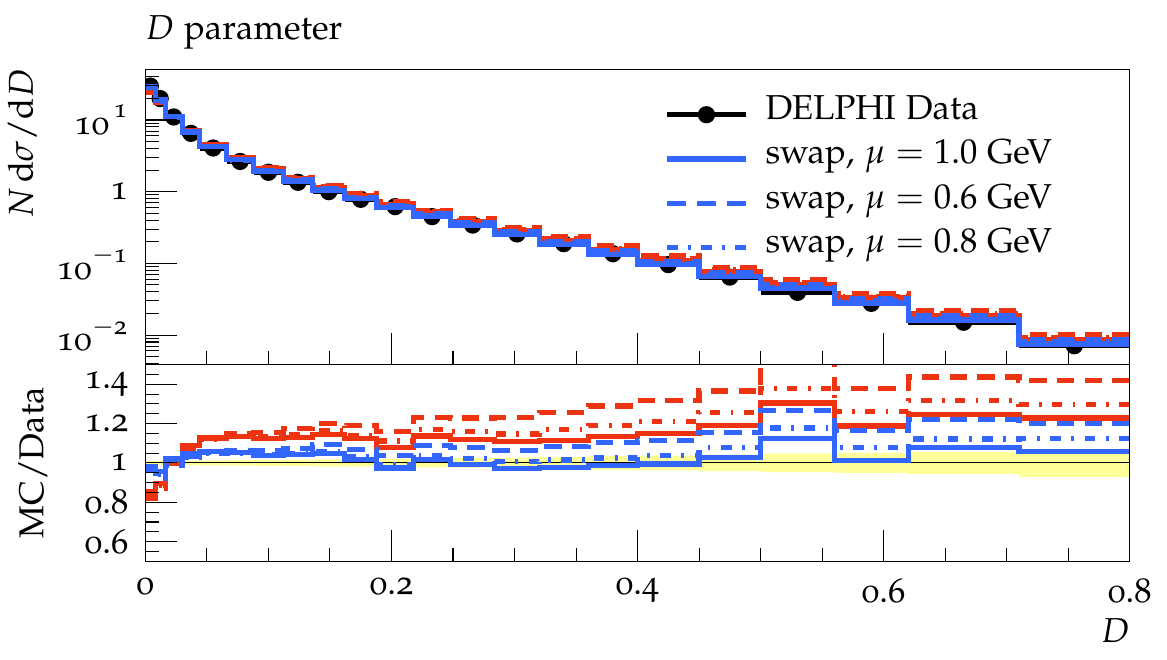}}
\scalebox{.735}{\includegraphics{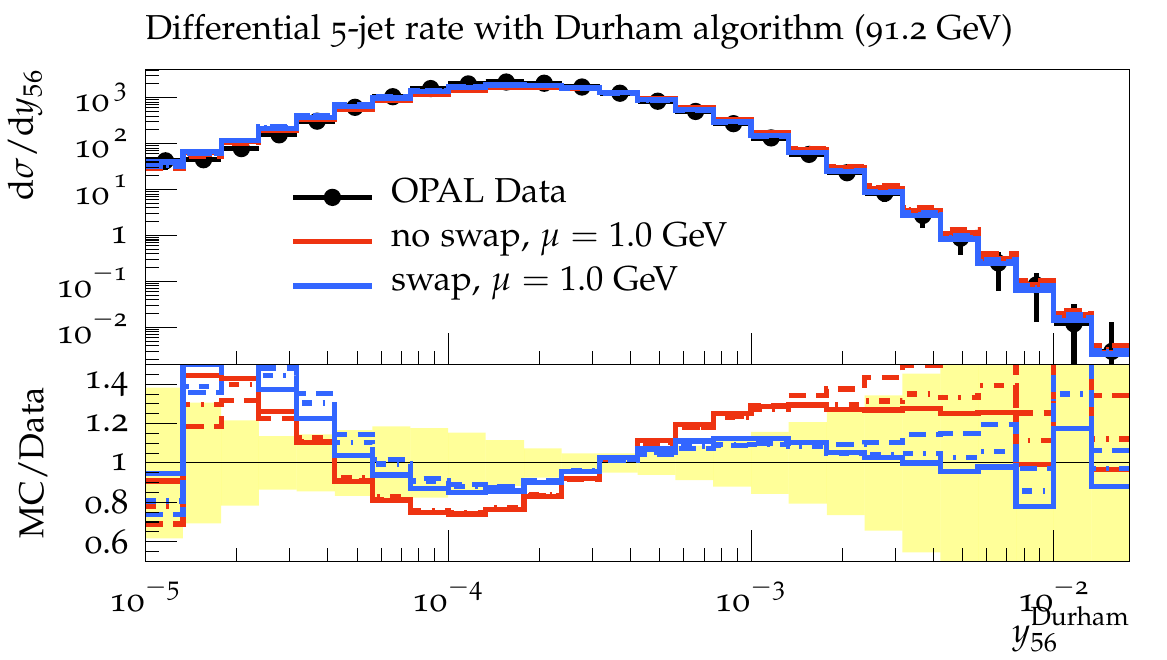}}
\caption{ C and D parameter as measured by \cite{Abreu:1996na}. Differential 5-jet 
rate as measured by \cite{Pfeifenschneider:1999rz}. For details see Sec.~\ref{sec:Results}.}
  \label{fig:LEP5}
\end{figure}

\vspace{-3mm}
In order not to bias\footnote{By assuming an improved coherence picture after rearranging
 color dipoles this statement might be questionable.} the results by tuning 
we choose to use the tuned values of the $\tilde{Q}$ shower of Herwig\cite{Reichelt:2017hts}. 
Further tuning of the shower with the modifications 
described in this work will improve the description of data but is also able to hide the effects 
due to rearranging the colours. Namely, parameters controlling the Cluster fission mechanism 
might allow having similar effects, as the number of particles can be reduced either by 
splitting clusters less often or, as in this approach, by reducing the average dipole sizes.
With the choice to use the value tuned for the $\tilde{Q}$ shower two parameters are free. 
The value of the strong coupling is $\alpha^{\MSbareq}_S(M_Z)=0.118$ and the IR
cutoff $\mu$ is varied by $0.6/0.8/1.0~\text{GeV}$. 
We convert the \MSbar\;  value of $\alpha_S$ to the MC scheme (CMW) \cite{CMW} 
with the appropriate factor. With these values we 
show\footnote{For analysing and plotting  Rivet \cite{Buckley:2010ar} was used.}
\begin{itemize}
   \item  the charge multiplicity as measured in \cite{Decamp:1991uz}, see Fig.~\ref{fig:LEP2}
   \item the heavy jet mass from \cite{Barate:1996fi},see Fig.~\ref{fig:LEP2}
   \item the C- and D- parameter as measured in \cite{Abreu:1996na}, see  Fig.~\ref{fig:LEP5}
   \item the five jet rate measured here \cite{Pfeifenschneider:1999rz}, see Fig.~\ref{fig:LEP5}
\end{itemize}   

We see large effects (up to $40~\%$ for standard LEP observables) and an overall improvement with respect to the standard dipole showering. It is notable that the observables
 shown here are sensitive to multiple emissions and we have checked that
observables sensitive to fewer emissions are in general not described worse.
Fully tuned results including $\chi^2$ comparisons as well as the extension to LHC physics 
will be discussed in future work.
\vspace{-6mm}
\section{Conclusion}
\vspace{-3mm}
In this paper we concentrated on the colour assignment in commonly used dipole-like 
parton showers. We then developed a method to assign a probability to the rearrangement
of colour dipoles. The method allows producing 'shorter' dipole chains if the shower falsely produces heavy chains by averaging the emission angle.
We finally show example comparison to data and see that not only the rearrangement 
can have effects of the order of up to $40~\%$ in standard observables but also  
by choosing an independent tune LEP data is better described by the procedure.
Various future projects including formal proofs, comparison to resummation 
and physics application are proposed.
\vspace{-6mm}
\section*{Acknowledgments}
\vspace{-3mm}
I would like to thank G. Gustafson, L. L{\"o}nnblad, M. Sj{\"o}dahl and 
T. Sj{\"o}strand for comments and joyful discussions in Lund. 
I thank all my collaborators on Herwig and especially S. Pl{\"a}tzer, P. Richardson 
and S. Webster for critical discussions and support.
This project has received funding from the European Research 
Council (ERC) under the European Union's Horizon 2020 
research and innovation programme, grant agreement No 668679. 

\vspace{-5mm}
\bibliography{colorea}

\end{document}